\documentclass[letterpaper, conference]{IEEEtran}
\usepackage{siunitx}
\usepackage{todonotes}
\ifCLASSINFOpdf
\else
\fi
\begin{document}
%
% paper title
% can use linebreaks \\ within to get better formatting as desired
\title{QUICker Connection Establishment\\ with Out-Of-Band Validation Tokens}

% author names and affiliations
% use a multiple column layout for up to three different
% affiliations
\author{\IEEEauthorblockN{Erik Sy\IEEEauthorrefmark{1}, Christian Burkert\IEEEauthorrefmark{1}, Tobias Mueller\IEEEauthorrefmark{1}, Hannes Federrath\IEEEauthorrefmark{1}, Mathias Fischer\IEEEauthorrefmark{1}}
\IEEEauthorblockA{\IEEEauthorrefmark{1}University of Hamburg, Germany\\ Email: \{sy, burkert, mueller, federrath, mfischer\}@informatik.uni-hamburg.de}
}

% conference papers do not typically use \thanks and this command
% is locked out in conference mode. If really needed, such as for
% the acknowledgment of grants, issue a \IEEEoverridecommandlockouts
% after \documentclass

% for over three affiliations, or if they all won't fit within the width
% of the page, use this alternative format:
% 
%\author{\IEEEauthorblockN{Michael Shell\IEEEauthorrefmark{1},
%Homer Simpson\IEEEauthorrefmark{2},
%James Kirk\IEEEauthorrefmark{3}, 
%Montgomery Scott\IEEEauthorrefmark{3} and
%Eldon Tyrell\IEEEauthorrefmark{4}}
%\IEEEauthorblockA{\IEEEauthorrefmark{1}School of Electrical and Computer Engineering\\
%Georgia Institute of Technology,
%Atlanta, Georgia 30332--0250\\ Email: see http://www.michaelshell.org/contact.html}
%\IEEEauthorblockA{\IEEEauthorrefmark{2}Twentieth Century Fox, Springfield, USA\\
%Email: homer@thesimpsons.com}
%\IEEEauthorblockA{\IEEEauthorrefmark{3}Starfleet Academy, San Francisco, California 96678-2391\\
%Telephone: (800) 555--1212, Fax: (888) 555--1212}
%\IEEEauthorblockA{\IEEEauthorrefmark{4}Tyrell Inc., 123 Replicant Street, Los Angeles, California 90210--4321}}

% use for special paper notices
%\IEEEspecialpapernotice{(Invited Paper)}

% make the title area
\maketitle

\begin{abstract}
QUIC is a secure transport protocol that improves the performance of HTTPS. An initial QUIC handshake that enforces a strict validation of the client's source address requires two round-trips. In this work, we extend QUIC's address validation mechanism by an out-of-band validation token to save one round-trip time during the initial handshake. The proposed token allows sharing an address validation between the QUIC server and trusted entities issuing these tokens. This saves a round-trip time for the address validation. Furthermore, we propose distribution mechanisms for these tokens using DNS resolvers and QUIC connections to other hostnames. Our proposal can save up to 50\% of the delay overhead of an initial QUIC handshake. Furthermore, our analytical results indicate that \SI{363.6}{ms} in total can be saved for all connections required to retrieve an average website, if a round-trip time of \SI{90}{ms} is assumed.
\end{abstract}
% IEEEtran.cls defaults to using nonbold math in the Abstract.
% This preserves the distinction between vectors and scalars. However,
% if the conference you are submitting to favors bold math in the abstract,
% then you can use LaTeX's standard command \boldmath at the very start
% of the abstract to achieve this. Many IEEE journals/conferences frown on
% math in the abstract anyway.

% no keywords

% For peer review papers, you can put extra information on the cover
% page as needed:
% \ifCLASSOPTIONpeerreview
% \begin{center} \bfseries EDICS Category: 3-BBND \end{center}
% \fi
%
% For peerreview papers, this IEEEtran command inserts a page break and
% creates the second title. It will be ignored for other modes.
\IEEEpeerreviewmaketitle

\section{Introduction}

This paper investigates the design of the QUIC protocol~\cite{ietf-quic-transport-19}, which is currently standardized. It is a secure transport protocol designed to replace TLS over TCP within the upcoming HTTP/3 version~\cite{ietf-quic-http-19}.
As the world wide web is closely tied to the Hypertext Transfer Protocol (HTTP) and the standardization work on QUIC receives widespread support, we expect the QUIC protocol to be widely deployed on the Internet within the next years.

The majority of web traffic consists of short-lived connections, for which the connection establishment represents a significant delay overhead~\cite{langley2017quic}.
QUIC's initial handshake requires two round trips to establish the connection. 
%QUIC reduces this overhead by merging the TCP and TLS handshakes, which require 1 1/2 and two handshakes, respectively, to a two round trip handshake. 
One round-trip accounts for the cryptographic connection establishment and the other for a challenge-response mechanism known as stateless retry, which validates the source address claimed by the client to prevent IP spoofing attacks.
Moreover, QUIC provides  zero round-trip time handshakes for resumed sessions. This allows clients to send encrypted requests directly without waiting for the server's first handshake messages.

% - Contribution
To further improve the performance of QUIC's initial handshake, we propose a mechanism to save one additional round trip by outsourcing the address validation mechanism.

% - Example
To illustrate the basic idea, we assume a website (google.com) that trusts a DNS resolver (Google DNS) to issue address validation tokens. Thus, if a client resolves google.com at Google DNS, it also retrieves a valid token for its source address. Subsequently, the client includes this out-of-band validation token in its connection request sent to google.com. Later, the web server validates that the presented token matches the claimed source address of the client. If so, the address validation is completed and in total, a round-trip time has been saved.
To put this into perspective, a typical round-trip time is below \SI{45}{ms} in North America and below \SI{90}{ms} for transatlantic connections~\cite{Verizon}.
Nonetheless, some regions in the world suffer from high network latencies, often exceeding \SI{300}{ms}~\cite{formoso2018deep}.
Thus, a saved round-trip time has a significant impact on the performance of the connection establishment.

In summary, this paper makes the following contributions:
\begin{itemize}
\item We propose out-of-band validation tokens that enable a shared address validation between a QUIC server and trusted entities issuing such tokens.

\item We propose mechanisms to distribute out-of-band validation tokens via DNS resolvers and other QUIC connections.

\item We demonstrate the performance improvements gained by out-of-band validation tokens. Our results indicate that our proposal saves up to 50\% of the delay overhead of initial QUIC connection establishments. Furthermore, the distribution of out-of-band tokens via DNS resolvers allows saving a round-trip time for almost all of the connections required to load an average website.

\end{itemize}

The remainder of this paper is structured as follows: Section~\ref{sec:Problem} introduces QUIC's stateless retry and describes the performance problems of QUIC's connection establishment that we aim to solve. 
Section~\ref{sec:Delegation} summarizes the proposed out-of-band validation token.
Evaluation results and a discussion of our proposal are presented in Section~\ref{sec:Evaluation}. 
Related work is reviewed in Section~\ref{sec:Related}.
Section~\ref{sec:Conclusion} concludes the paper.

\section{Problem Statement} \label{sec:Problem}

In this section, we review the stateless retry mechanism as known from IETF QUIC~\cite{ietf-quic-transport-19}.
Subsequently, we introduce the performance problem of QUIC's source address validation that we aim to solve.

\subsection{Stateless retry} \label{sec:stateless_retry}

QUIC servers can optionally include a challenge-response mechanism in the handshake to validate the client's source address before proceeding with the cryptographic connection establishment.
In the following, we first describe the protocol flow of this mechanism, which is known as a stateless retry within the QUIC terminology.
Subsequently, we present details on the generation of address validation tokens.

\paragraph{Protocol flow}

\begin{figure}
\centering
\includegraphics[width=0.47 \textwidth]{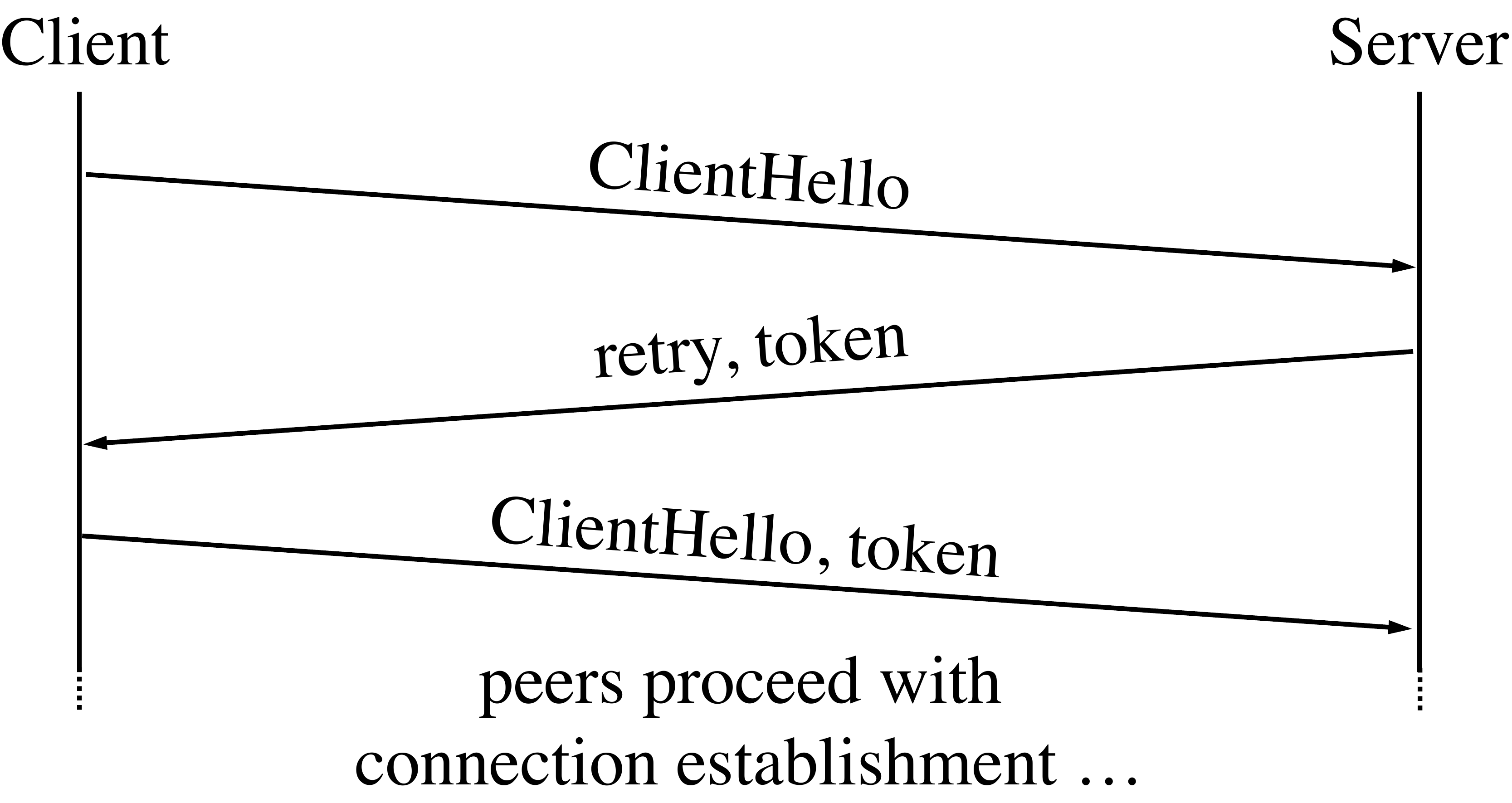}
\caption{Schematic of QUIC's stateless retry during an initial connection establishment. }
\label{fig:Quic_retry}
\end{figure}

Figure~\ref{fig:Quic_retry} provides an overview of a stateless retry during a client's connection attempt.
The client starts the connection attempt by sending a ClientHello message.
In case of a stateless retry, the server responds with a retry packet that contains an address validation token.
Upon receiving the server's response, the client must repeat the received token when resending its ClientHello message.
This mechanism allows the server to validate the client's source address.
However, this mechanism increases the delay of the connection establishment by a round-trip time.
Note, that a server can abort the connection establishment if, during a stateless retry, a received token does not validate the claimed IP address.

\paragraph{Token generation}\label{sec:token_gen}

The draft of the QUIC protocol does not suggest a specific mechanism to implement the generation of tokens because the server creating this token is also consuming it.
These tokens should be difficult to guess and should not allow a network observer to link several QUIC connections to the same client~\cite{sy2019quic}.

For example, these tokens can be constructed using a Hashed Message Authentication Code (HMAC) based on SHA-256~\cite{rfc6234}.
Basically, the HMAC function depends on a secret key and the client's IP address, while the token is the (truncated) HMAC value.
To avoid that the same token is issued repeatedly, the client's IP address can be concatenated with a cryptographic nonce in the HMAC function.
However, to successfully validate such HMAC values, the used nonce must be encoded in the token presented by clients.
Furthermore, an identifier for the used secret key can be appended to the token to facilitate key management.
Note, that the computation of HMACs requires around 10 CPU cycles per byte of input data~\cite{benchmark}.
Thus, off-the-shelf server CPU's can usually construct several ten thousands or even hundred thousand validation tokens per second.
 %
%Thus, the construction of these tokens might use the Advanced Encryption Standard~(AES) with a secret key length of 128~bits (16-byte) deploying an authenticated encryption mode such as Offset Codebook (OCB) or Galois/Counter Mode (GCM).
%The plaintext may consist just of the client's IP address and a cryptographic nonce is used as the initialization vector (IV).
%As clients have no possession of the secret key, these tokens present an opaque data block for them.
%Furthermore, it is not feasible for clients to generate valid tokens for arbitrary source addresses.

This mechanism of source address validation is called stateless because the server is not required to keep state about issued tokens but only needs to know the algorithm for the token creation and the secret key to validate a presented token.
By sharing this secret key among a group of servers, a validation of token can be performed by a member of this group that did not issue the token itself.
For example, a stateless retry can request the client to conduct the handshake with another server instance for the purpose of load-balancing.
In this case, the other server needs the used secret key to validate that the presented token matches the claimed source address. 

\paragraph{Address validation for future connections}
To allow a strict address validation without causing a retry and an additional round-trip, the QUIC server can issue tokens via the \textit{new\_token} frame over an already established connection.
The client caches this token for use in future connections.
If the client wishes to establish a new connection to the same server, it includes the cached token within its initial message.
Upon receiving the client's initial message, the server validates that the presented token matches the claimed source address.
If this is the case, the server accepts the claimed source address as validated and proceeds with the cryptographic connection establishment.
In total, this practice saves a round-trip time compared to the source address validation using a stateless retry. 

\subsection{Performance limitations of QUIC's address validation}

A stateless retry increases the delay overhead of the connection establishment by a round-trip time.
Validation tokens for future connections solve this problem for all revisits to the same hostname.
However, these tokens are not available for the first connection establishment to a specific hostname.
Furthermore, a presented validation token for future connections can be invalid if it expired or the client's source address (as seen by the server) changed in the meantime.
Upon receiving such an invalid token, the server responds with a stateless retry if the address validation is required before proceeding with the cryptographic connection establishment.
Thus, there are several situations in which a stateless retry is likely to occur during the establishment of a connection.
Each time this leads to a performance penalty of a round-trip time.

Note, that on average the retrieval of a website requires about 20~connections to different hostnames~\cite{sy2019enhanced}.
This indicates, that web browsing causes a large number of short-lived connections for which the connection establishment can present a significant overhead.
 Furthermore, websites on average require four sequential connection establishment for their retrieval because downloaded resources often trigger new connections to different hostnames~\cite{sy2019enhanced}.
Assuming, that the triggering and triggered connections saved a round-trip during their address validation, then the loading of a website can save more than a round-trip time to complete.  

%\subsection{Threat model}\label{sec:threat}
%
%This section defines our threat model to clarify the security considerations of the presented problem.
%
%We assume that the generated validation tokens are cryptographically secure.
%As a result, it is not feasible for the considered adversary to generate a valid token for an arbitrary source address.
%However, the attacker can spoof the source IP address of its outgoing packets.
%
%The considered security threat affects the objective of availability.
%Hereby, the adversary can attempt to directly exhaust the resources of a QUIC server by requesting connections from several spoofed source addresses.
%This eventually leads to a denial-of-service attack if the server spends too many resources on the connection requests from clients claiming unvalidated source addresses.
%
%Furthermore, the adversary can also spoof the source address of a victim's endpoint.
%If so, the QUIC server will send its response to the victim's source address, which is known as a reflection attack.
%In this scenario, the adversary aims to saturate the bandwidth available to the victim's endpoint.
%Note, that the server response can be larger than the adversary's initial packets.
%Therefore, this approach may amplify the data volume of the attack on the victim's availability compared to a direct attack by the adversary on the victim's endpoint.

\section{Out-of-band validation token} \label{sec:Delegation}

This section introduces the out-of-band validation token for the QUIC protocol.
Subsequently, distribution mechanisms for such out-of-band tokens are proposed using DNS resolvers and QUIC connections to other hostnames.

\subsection{Token design}

Address validation tokens present a defense mechanism against source address spoofing by malicious clients.
For this purpose, the QUIC server compares the claimed client address with the previously observed source address encoded in the presented token.
So far, only QUIC servers themselves can issue address validation tokens for their connections.
Out-of-band validation tokens extend this mechanism by allowing external entities to issue these tokens.

The generation of these token follows a similar approach as described in Section~\ref{sec:stateless_retry}.
Thus, the QUIC server is required to share instructions and a secret key with the corresponding external entity, that allow the generation of valid out-of-band tokens for the client's source address.
Upon receiving an out-of-band token, the client imports it in its cache, marks it as received by an external entity, and associates the QUIC server's hostname to it.
To establish a fresh connection to the respective hostname, the client includes a cached token in the send initial packet.
If the client's cache contains several tokens, the client must prefer the usage of validation tokens received by the QUIC server itself over cached out-of-band tokens. 

The server may share different secret keys with different external entities.
This approach allows a selected invalidation of tokens that have been issued using a specific secret key.
Thus, the QUIC server can revoke the secret key provided to an external entity if, e.g., a large number of unrequited connection requests is observed that use tokens issued by the same key.
If a setup with dedicated secret keys per external entity is deployed, it is recommended to attach an identifier to the token, that indicates which key was used to generate the specific token.

Note, that according to the draft of IETF QUIC~\cite{ietf-quic-transport-19} the server treats an invalid token as if the client did not present a token.
Thus, the number of required round-trips during the connection establishment is identical if the client presents an invalid out-of-band token or the client's connection request does not contain a token at all.

\subsection{Token distribution mechanisms}

To substantiate the real-world benefit of out-of-band tokens, we present in this section two distribution mechanisms for such tokens.
First, we introduce the distribution via the Domain Name System (DNS).
Then, we describe a distribution mechanism using QUIC connections to other hostnames for this purpose.

\paragraph{Distribution via DNS resolver}

To save a round-trip time via the proposed out-of-band tokens, the client needs to receive the token before sending the connection request to the corresponding QUIC server.
Furthermore, clients query a domain name to look up the source address before they send their connection request.
Thus, DNS seems to be a suitable place to distribute out-of-band tokens as the connection request often directly follows the corresponding DNS lookup.
\begin{figure}
\centering
\includegraphics[width=0.47 \textwidth]{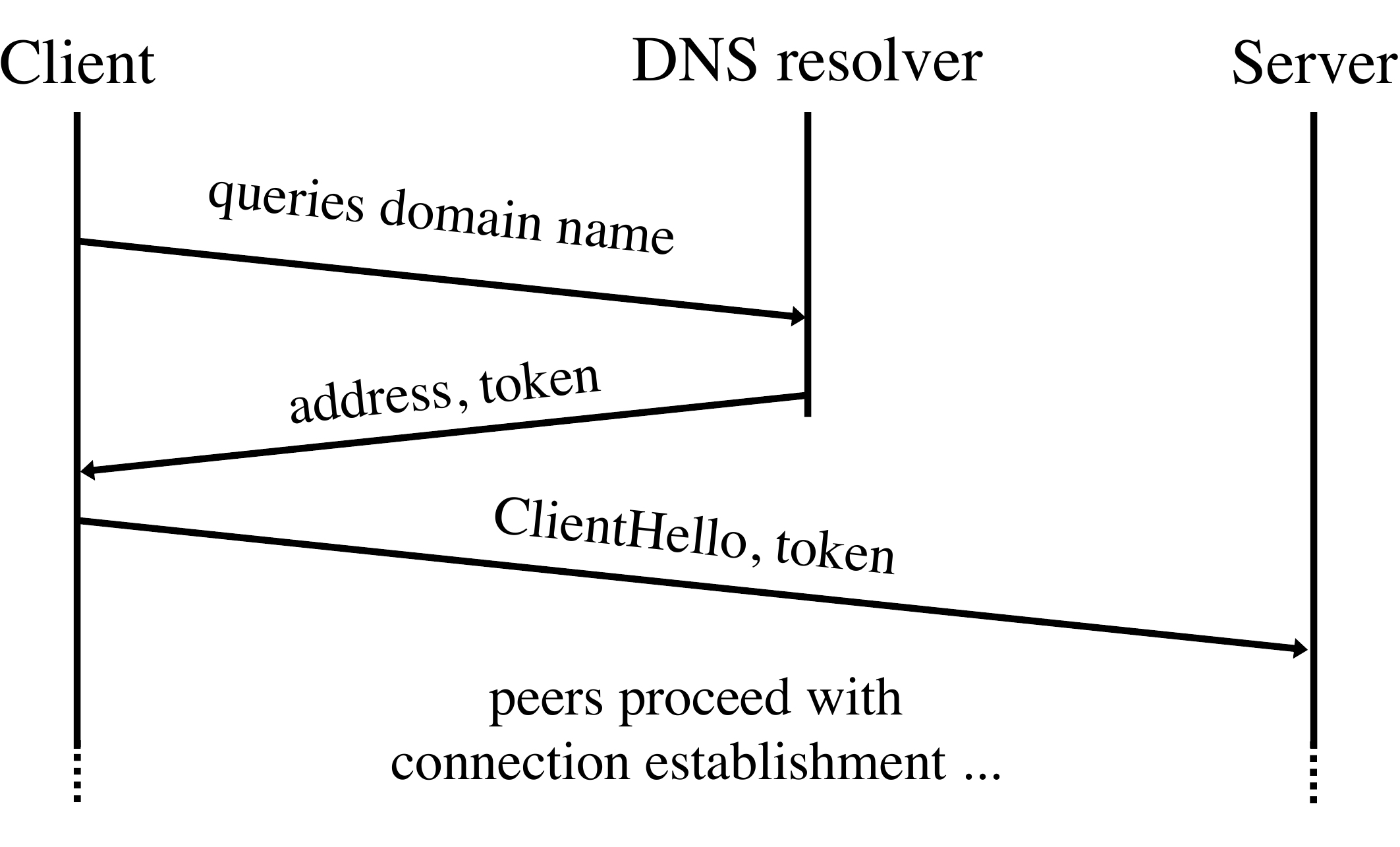}
\caption{Proposed connection establishment avoiding stateless retry by using a DNS resolver to issue validation token.}
\label{fig:DNS_query}
\end{figure}
Figure~\ref{fig:DNS_query} provides a schematic of this proposed distribution mechanism.
This proposal is not limited to a specific DNS standard and can be applied to the traditional DNS~\cite{rfc1035} and newer versions deploying transport encryption such as DNS over Transport Layer Security (DOT)~\cite{rfc7858} and DNS over HTTPS (DOH)~\cite{rfc8484}.
In general, the DNS protocol needs to be extended by a new record type, which we define as QUICTOKEN\@.

If the client wishes to establish a fresh QUIC connection to a domain name for which it has not a cached token for future connections available, it queries the domain name (including the QUICTOKEN record type) as shown in Figure~\ref{fig:DNS_query}.
The DNS resolver proceeds with the default resolution of the source address associated with the domain name.
Additionally, if the DNS resolver supports the record type QUICTOKEN and is capable to generate valid out-of-band tokens for this queried domain name, it can include such a token in the response sent to the client.

Note, that to construct valid out-of-band tokens, the resolver needs to be trusted by the server hosting the specific domain name.
Thus, the respective server operator shared in advance the instructions and the secret keys required to generate valid tokens for this domain.

Upon receiving the source address and the token, the client constructs its QUIC connection request and attach the obtained out-of-band token to it before sending it to the received server source address. 
Subsequently, the server validates the presented token and proceeds with its normal connection establishment.

To ensure that clients do not reuse tokens across different connections, it is required that the record type QUICTOKEN must not be cached except by the client.
This can be realized by setting the Time to Live (TTL) of the QUICTOKEN record type to zero seconds.
Note, that the DNS specification explicitly allows TTL of zero seconds~\cite{rfc2181}.
Furthermore, that the DNS specification~\cite{rfc2181} allows each record type to have its own TTL.
As a result, this configuration of QUICTOKEN does not affect the caching mechanisms of for example A or AAAA record types.

A limitation of this distribution mechanism arises if the DNS resolver is located within the same private network as the client.
In this case, the client's source address as seen by the DNS resolver might mismatch the publicly visible source address as seen by the QUIC server.
Thus, the address validation is likely to fail because the source address encrypted in the token does not match the claimed source address as observed by the QUIC server.
This issue can be solved by for example moving the DNS resolver to a public IP address or using STUN~\cite{rfc5389} to learn the client's public source address.

The computational overhead introduced to the DNS resolver when constructing out-of-band validation tokens presents another limitation of our proposal.
Large connection-oriented DNS can have about 24K active connections and serve up to 230k queries per second~\cite{zhu2015connection}.
Thus, it seems beneficial to use a lightweight mechanism for constructing these tokens such as the discussed HMAC functions (see Section~\ref{sec:stateless_retry}\,b)).
However, widespread mechanisms such as EDNS Client Subnet~\cite{rfc7871} or Round-robin DNS~\cite{rfc1794} also cause an increased overhead for DNS resolvers to realize performance optimizations.

\paragraph{Distribution via other QUIC connections}

\begin{figure}
\centering
\includegraphics[width=0.47 \textwidth]{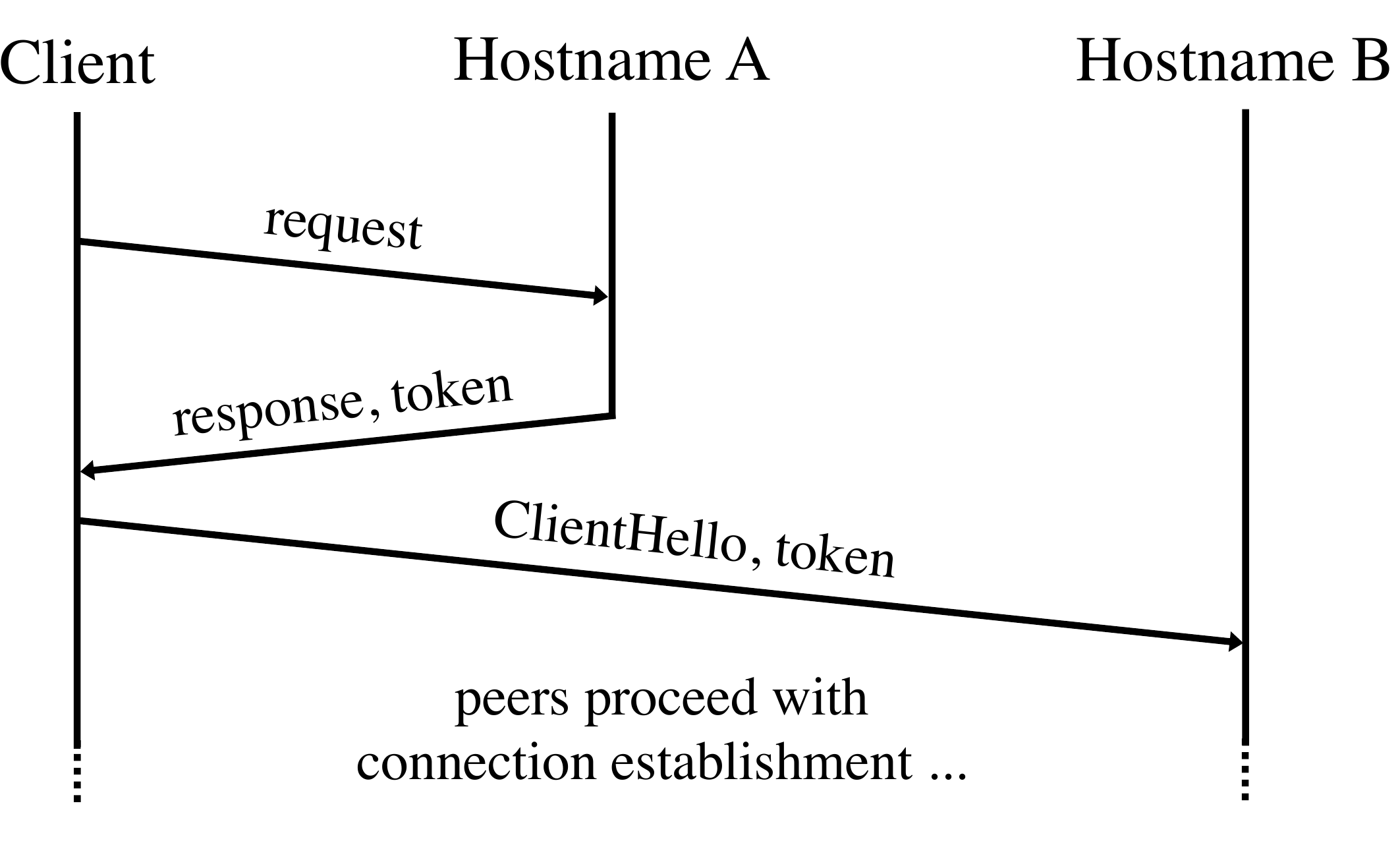}
\caption{Proposed distribution mechanism using the QUIC connection to hostname~A to receive an out-of-band token valid for hostname~B.}
\label{fig:Quic_query}
\end{figure}

This distribution mechanism assumes that the client first establishes a QUIC connection to hostname~A before it sends a connection request to hostname~B.
Furthermore, we assume that hostname B allows hostname A to issue valid out-of-band tokens for its service and therefore shares instructions and its secret key required to construct these tokens for arbitrary source addresses.
We propose a new EXTERNAL\_TOKEN frame for the QUIC protocol, which allows QUIC servers to provide clients with out-of-band tokens for arbitrary hostnames.
However, tokens for future connections to the same hostname~A should use the existing NEW\_TOKEN frame.
Note, that tokens for future connections are regarded as trustworthy as they are issued by the same server which is also consuming them.
However, out-of-band tokens are not treated as trustworthy because the client does not validate that the entity issuing these tokens is authorized to do so.
Compared to the NEW\_TOKEN frame of the QUIC protocol, tokens received via the EXTERNAL\_TOKEN frame are only used to establish a fresh connection if the client would otherwise send the connection request without an attached address validation token.

Figure~\ref{fig:Quic_query} shows a schematic of this distribution mechanism.
The client has an established QUIC connection to hostname~A.
Hostname~A reasons based upon its provided response, that the client is likely to establish a connection to hostname~B.
This is for example the case, if the provided response contains a hyperlink to a resource hosted by hostname~B.
To speed up this connection establishment between the client and hostname~B, hostname~A decides to provide an out-of-band token for the client's source address valid for hostname~B.

Upon receiving this EXTERNAL\_TOKEN frame from hostname~A, the client checks first if it has a token for future connections for hostname~B.
If not, the client establishes a fresh connection to hostnames~B by attaching the received out-of-band token to its connection request.
Otherwise, the client will prefer to include its cached token for future connections (received in a previous connection to the hostname~B) in the connection request to hostname~B.
Upon receiving the client's connection request, hostname~B validates the included address validation token and proceeds with the usual connection establishment.

It seems reasonable to expect that a QUIC server will only issue out-of-band tokens for other hostnames for which it is likely that the client will soon connect to them.
Out-of-band tokens should have an expiration mechanism, thus received tokens may expire if no connection is established to a corresponding hostname within a short period.

\section{Evaluation and discussion} \label{sec:Evaluation}

To demonstrate the feasibility of our proposal, we evaluate and discuss aspects of its performance, security, privacy, and scalability. 

\subsection{Performance}

In this section, we present a performance evaluation for out-of-band tokens.
First, we describe our results with respect to the establishment of a single QUIC connection.
Subsequently, we evaluate the performance impact of our proposal on an average website visit.

\paragraph{Benefits for single connections}

\begin{figure}
\centering
\includegraphics[width=0.47 \textwidth]{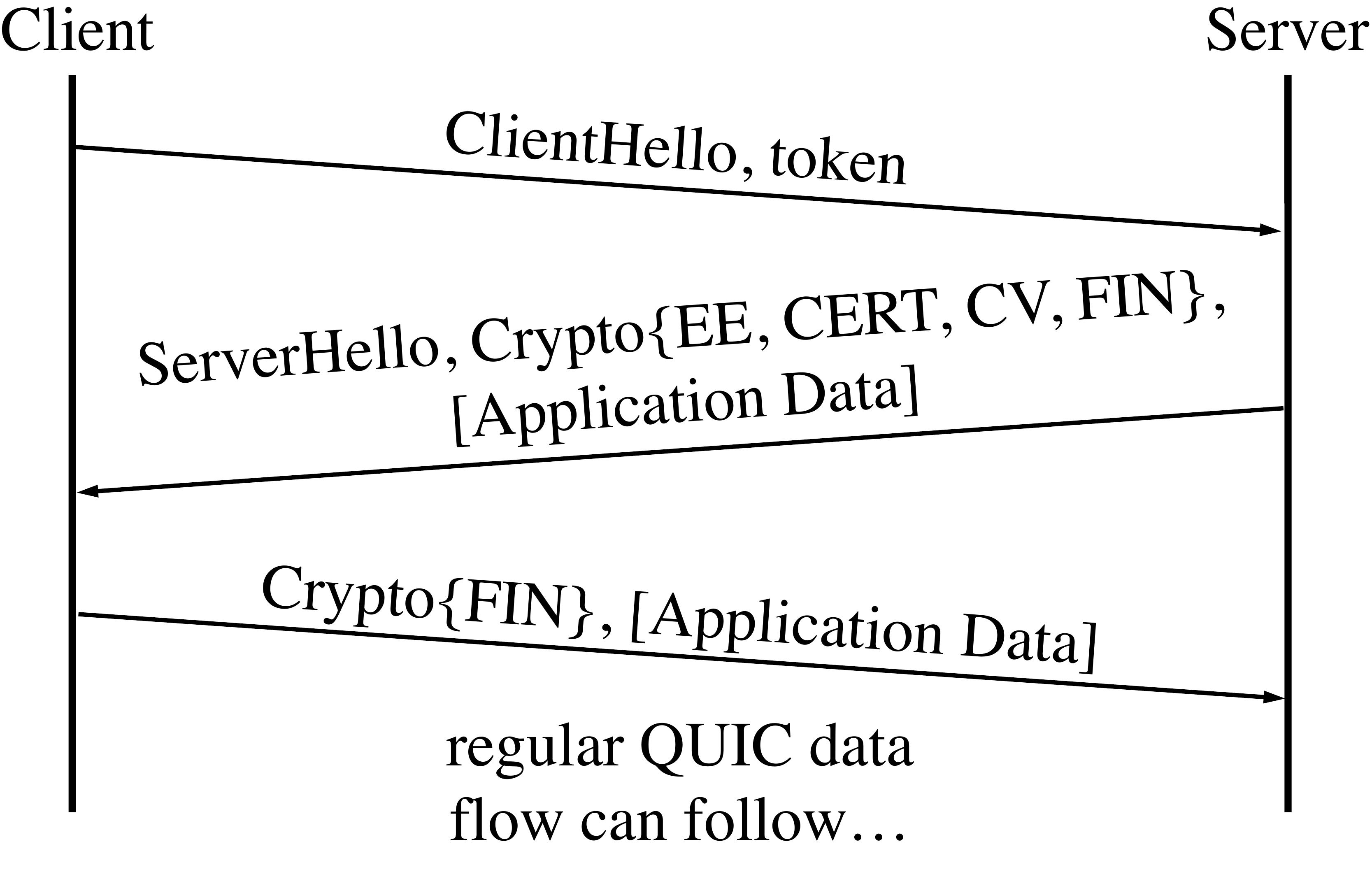}
\caption{Initial QUIC connection establishment using previously retrieved out-of-band validation token.}
\label{fig:QUIC_reject}
\end{figure}

Using an out-of-band token to validate the client's source address saves a round-trip compared to using a stateless retry.
Figure~\ref{fig:QUIC_reject} shows QUIC's initial handshake where the client presents an out-of-band token within the initial packets sent to the server.
In case of a valid token, the server directly proceeds with the cryptographic handshake by sending its ServerHello message.
The cryptographic handshake follows the TLS~1.3 protocol.
Thus, the server uses transport encryption to transmit its Encrypted Extensions (EE), Certificate (CERT), Certificate Verify (CV) and Handshake Finished (FIN) messages.
In total, it requires only a single round-trip from the ClientHello until the client transmits its own FIN message and is ready to send encrypted application data.
As the QUIC protocol is still work in progress, only experimental implementations of its design exist.
Thus, we will use an analytical model to approximate the performance benefit of our proposal on the delay overhead of the connection establishment.
For this evaluation, we approximate the delay overhead for the initial connection establishment as shown in Equations~\ref{eq:default} and~\ref{eq:proposal}.
Here, $t_\mathit{Default}$ and $t_\mathit{Proposal}$ indicate the delay overhead for the current status quo and our proposal, respectively.  
Furthermore, RTT denotes the round-trip time between both peers and $t_{proc}$ marks the total time required by the peers to process the connection establishment.
\begin{equation}\label{eq:default}
  t_\mathit{Default}(RTT) = t_{proc}  + 2 * RTT
\end{equation}
\begin{equation}\label{eq:proposal}
  t_\mathit{Proposal}(RTT) = t_{proc} + RTT
\end{equation}
Within our analytical model, we assume that the processing of the connection setup $t_{proc}$ takes \SI{40}{ms} independently of the round-trip time.
We chose \SI{40}{ms} because this approximates the time by the TLS~1.3 over TCP protocol stack for a similar task~\cite{sy2019enhanced}.

\begin{figure}
\centering
\includegraphics[width=0.47 \textwidth]{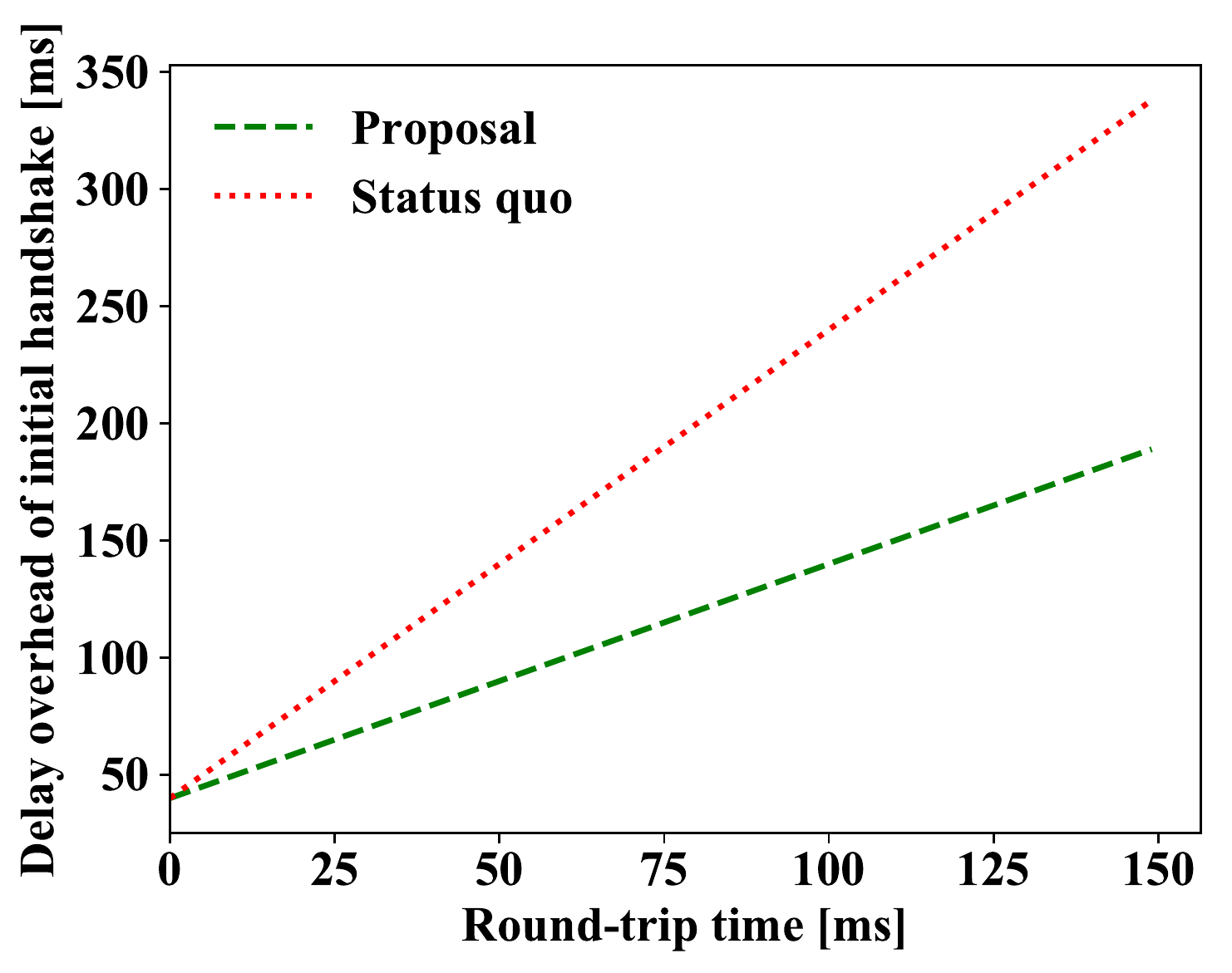}
\caption{This plot shows the simulated delay overhead of QUIC's initial connection establishment over the round-trip time. The red, dotted line plots QUIC's default behavior and the green, dashed line deploys our proposed out-of-band validation tokens.}
\label{fig:Plot_single_connection}
\end{figure}

Figure~\ref{fig:Plot_single_connection} plots the delay overhead of the initial handshake over the round-trip time for our analytical model.
We find, that the green, dashed line indicating our proposal provides significantly better results than QUIC's status quo marked by a red, dotted plot.
The performance improvement achieved by our proposal depends on the RTT\@.
Assuming a  transatlantic connections with a round-trip time of \SI{90}{ms}~\cite{Verizon}, we find that a connection establishment using our proposal requires only 60\% of the default delay overhead.  
Furthermore, we derive from Equation~\ref{eq:default} and~\ref{eq:proposal} that our proposal reduces the investigated delay overhead by 50\% when RTT converges to infinity. 

\paragraph{Gains for web browsing}

As the QUIC protocol will be a building block of the upcoming HTTP/3 network protocol, it seems interesting to evaluate the performance impact of our proposal on website retrieval.
A recent study reported, that the retrieval of popular websites requires on average 20.24~encrypted connections to different hostnames~\cite{sy2019enhanced}.
For this evaluation, we assume that all of these hostnames support the QUIC protocol and that they all enable the client's DNS resolver to issue out-of-band tokens.
Thus, we find that we can save a round-trip time during each connection establishment if the corresponding web server enforces a strict source address validation before proceeding with the cryptographic handshake.

Furthermore, it is found by~\cite{sy2019enhanced}, that the average popular website requires up to 4.04~sequentially established connections.
This finding can be attributed to the fact, that a retrieved web resource often triggers the establishment of additional connections to retrieve further resources.
Thus, saving a round-trip time via the proposed out-of-band tokens allows in total to save on average 4.04~round-trips until all required connections are established.
Assuming a round-trip time of \SI{90}{ms}, as it is typical for transatlantic connections~\cite{Verizon},
we find that \SI{363.6}{ms} can be saved until the last connection for retrieving the website is established.

\subsection{Security}

In this section, we review possible security concerns with respect to out-of-band validation tokens.
First, we address the impact of our proposal on the mitigation of Denial-of-Service attacks.
Then, we look at risks arising from using address validation tokens from possibly unauthorized origins.

\paragraph{Denial-of-Service attacks}
By sharing the instructions and the secret keys to generate address validation tokens with other entities, the risk that this confidential information gets compromised increases.
In case of a compromise, an adversary can issue tokens for arbitrary source addresses.
Thus, the adversary can send connection requests with a spoofed source address to the QUIC server, that contain a valid token for the claimed address.
As a result, the server experiences a large number of spoofed connection requests that consume its available resources up to a Denial-of-Service attack.
To mitigate such an event, the server should monitor connection requesrs associated with trusted secret keys.
If the number of spoofed connection requests exceeds a threshold, the server revokes that specific secret key to mitigate Denial-of-Service attacks.
After such a revocation all tokens issued by this key are treated as invalid.
However, the revocation of a secret key might also cause a stateless retry for legitimate connection requests and thus causes a performance degradation for these connection attempts.
To address this security versus performance tradeoff, it is advised to provide different secret keys to different entities.
Thus, a revoked key affects only validation tokens expected to be issued by a specific entity.
In total, key revocation provides an effective mechanism to protect against the considered Denial-of-Service attacks.

\paragraph{Unauthorized origins}
In our proposal, the client does not validate whether an external entity is authorized by the affected QUIC server to issue out-of-band tokens.
Thus, it is necessary to review the case in which an unauthorized external entity issues invalid out-of-band tokens.
The draft of IETF QUIC~\cite{ietf-quic-transport-19} instructs that servers treat invalid tokens (for future connections) as if the client did not present a token at all.
Therefore, the client experiences no drawbacks by presenting an invalid out-of-band validation token and in total, the client does achieve the same performance as including no token in the connection request.
However, if the client has a valid token for future connections and an invalid out-of-band token from an unauthorized origin in its cache, then only a connection request including the token for future connections can save a round-trip during the address validation.
Concluding, tokens for future connections are more trustworthy, as they have been retrieved via an authenticated connection to the respective QUIC server.
For this reason, the usage of tokens for future connections should be preferred over out-of-band validation tokens because clients do not validate that a trust-relation  between the entity issuing the out-of-band tokens and the QUIC server consuming them exists.
However, out-of-band tokens are preferential compared to using no token at all, as they can reduce the delay overhead of the connection establishment by up to a round-trip time but always achieve at least the performance of a handshake without a token. 

\subsection{Privacy}

QUIC's address validation tokens can be used to link the connection in which the token was issued to the future connection in which the same token is presented by the client~\cite{sy2019quic}.
The proposed out-of-band token allows the same correlation across both connections in which the corresponding token is exchanged.
However, in the case of out-of-band tokens, the entity issuing the token might differ from the one to whom it is presented to during the connection request.
Thus, the privacy aspects of the proposed token are different from the known address validation token for future connections as it allows linking user activities across different entities.
To evaluate the privacy impact of our proposal, we first investigate the distribution of tokens via DNS resolvers.
Subsequently, we review the issuing of out-of-band tokens via QUIC connections to other hostnames.

\paragraph{Distribution via DNS resolvers}

First of all, we find that the described correlations of a client's connections are only feasible if the involved entities collude with each other with the aim to track a client's online activities.
We note, that a collusion between a DNS resolver and a QUIC server does provide significant opportunities to identify the same client across these services, e.g.,
the close temporal proximity between DNS lookups and subsequent connection requests to QUIC servers.
We conclude, that a collusion between both services can already reduce the client's anonymity set based on the timing of the corresponding requests.
Moreover, the client's source address will usually be the same when communicating with the DNS resolver and the QUIC server, which further facilitates the linking of the user's activities.
Furthermore, the DNS resolver can respond with a unique server source address upon the client's DNS query, which is especially feasible for IPv6 addresses.
Subsequently, both services can use this address to link the user across their services. 
Possibly, further opportunities to link users arise from the usage of DNS record types other than address records (A or AAAA) such as entries for Encrypted Server Name Indication (ESNI) for TLS~1.3~\cite{ESNI}.
In total, it seems not feasible to prevent user tracking between DNS provider and QUIC server operators if these entities collude with each other.

\paragraph{Distribution via other QUIC connections}

Similar to the DNS-based scenario, several operators of QUIC servers can share their clients' source addresses and the time of the requests to match user profiles across services.
However, web applications are usually capable of triggering a request to another URL using a HTTP redirect or hyperlink.
This allows the colluding entities to encode a client identifier within the used URL\@.
Thus, if the client follows the received HTTP redirect or hyperlink, both colluding entities can share the respective client's profile based on this unique client identifier. 
Concluding, it does not seem to be feasible to prevent user tracking across colluding QUIC servers in a real-world context.

To mitigate the privacy impact of our proposal, we recommend expiring out-of-band tokens within short periods.
As proposed in~\cite{sy2018tracking}, ten minutes seem to be a reasonable limit for comparable mechanisms that enable user tracking.

\subsection{Scalability}

The proposed distribution mechanisms require the establishment of trust-relations between different hostnames or even services.
Large corporations such as Google or Cloudflare that cover several thousands of websites and provide their own popular DNS resolvers can easily deploy our proposal for their own services.
Furthermore, it seems feasible that large Internet corporations establish the required trust between each other based on personal contacts to allow issuing out-of-band tokens across their services.
As a result, this practice would only benefit the client's connection establishment with these few online services, while most online services do not significantly benefit from the performance improvements achieved via out-of-band tokens. 

To make this proposal available to every web service, it requires an automated approach to establish the required trust-relations and subsequently share, update and possibly revoke the secret keys required for issuing out-of-band tokens.
Possibly, it requires a trusted entity similar to the CA/Browser Forum which can provide a whitelist of trustworthy DNS resolvers.
This whitelist can be used by QUIC servers to share their secret keys required for issuing out-band-tokens with these DNS resolvers.

With respect to out-of-band tokens issued by other QUIC servers, a deviation of the ACME protocol~\cite{rfc8555} seems plausible to automate the process of establishing trust and conducting the required key management.
Here, the QUIC server first validates the legitimate interest of another QUIC server to issue out-of-band tokens.
Legitimate interest can be argued if the server, which intends to issue out-of-band tokens serves hyperlinks or HTTP redirects to the corresponding server, that consumes the out-of-band validation tokens.
If the interest is valid, both servers will subsequently exchange the required key material to issue such tokens.

To the best of our knowledge, no protocols suitable for these tasks exist.
Thus, we hope that this brief discussion of the scalability problem at hand fosters further research and development on the design of such protocols, that makes out-of-band validation tokens available to every web service.

\section{Related work}\label{sec:Related}

Performance improvements of the QUIC protocol with respect to the performance penalty caused by a stateless retry are actively discussed within the Internet Engineering Task Force (IETF) QUIC working group.
So far, these discussions focus on extending the number of entities that are allowed to issue address validation tokens for other hostnames either based on existing TLS trust-relations~\cite{sy2019quicker} or based on the source address from which a respective hostname is served~\cite{kazuho-quic-address-bound-token-00}.
Both prior contributions have limited applicability to avoid stateless retries.

\cite{sy2019quicker} does not mitigate a stateless retry during the first connection to a member of a group of hostnames that have an exiting trust-relation with each other.
Furthermore, these groups are usually rather small~\cite{sy2019enhanced}, thus only about 60\% of connections established during the first visit of an average website can benefit from this approach.
However, validation tokens obtained from a member of such a group can be considered as trustworthy because these members share secret cryptographic state with each other such as  a private key of a X.509 certificate.

\cite{kazuho-quic-address-bound-token-00} proposes to bind validation tokens to the address of the server, similar to the approach of the TCP Fast Open protocol~\cite{rfc7413}.
This approach is limited as it does not mitigate a stateless retry upon the initial connection establishment to a specific server source address and performance gains can only be realized on subsequent connections to a hostname served from the same source address.
Furthermore, the feature of connection reuse in HTTP/2~\cite{rfc7540} allows using an established connection to a server at a specific source address to request resources for another virtual host on the same server.
This feature of HTTP reduces the chance that a client requests another connection to the same server at the same source address.
Thus, it remains so far unclear to which extend this proposal improves the status quo.

This work extends the applicability of the discussed related work because clients can use out-of-band tokens upon the first connection request to any QUIC server, assuming that their DNS resolver is capable to provide a corresponding token.
Thus, this proposal outperforms the related work by saving up to 100\% of the stateless retries usually required if a strict source address validation is enforced.
To the best of our knowledge, this work is the first to propose the distribution of address validation tokens via DNS\@.

\section{Conclusions}\label{sec:Conclusion}

This paper proposes out-of-band validation tokens for a shared address validation between a QUIC server and trusted entities issuing these tokens.
Our evaluation indicates, that the proposed tokens enable significant performance gains for clients and servers without affecting the user's privacy and communication security.

\IEEEtriggeratref{21}
% can use a bibliography generated by BibTeX as a .bbl file
% BibTeX documentation can be easily obtained at:
% http://www.ctan.org/tex-archive/biblio/bibtex/contrib/doc/
% The IEEEtran BibTeX style support page is at:
% http://www.michaelshell.org/tex/ieeetran/bibtex/
\bibliographystyle{IEEEtranS}
% argument is your BibTeX string definitions and bibliography database(s)
\bibliography{IEEEabrv,sample-bibliography}
%
% <OR> manually copy in the resultant .bbl file
% set second argument of \begin to the number of references
% (used to reserve space for the reference number labels box)
%\begin{thebibliography}{1}

%\bibitem{IEEEhowto:kopka}
%H.~Kopka and P.~W. Daly, \emph{A Guide to \LaTeX}, 3rd~ed.\hskip 1em plus
%  0.5em minus 0.4em\relax Harlow, England: Addison-Wesley, 1999.

%\end{thebibliography}

% that's all folks
\end{document}